\documentclass[conference]{IEEEtran}
\IEEEoverridecommandlockouts

\usepackage{cite}
\usepackage{amsmath,amssymb,amsfonts}
\usepackage{algorithmic}
\usepackage{graphicx}
\usepackage{textcomp}
\usepackage{amsthm}
\newtheoremstyle{BoldHead}
  {0pt}{0pt}
  {\normalfont}
  {}
  {\bfseries}
  {.}
  {0.5em}
  {} 
\allowdisplaybreaks
\theoremstyle{BoldHead}
\newtheorem{theorem}{Theorem}
\newtheorem{lemma}{Lemma}
\newtheorem{remark}{Remark}
\newtheorem{definition}{Definition}
\newtheorem{example}{Example}

\usepackage{xcolor}

\newcommand{\msf}[1]{\mathsf{#1}}
\usepackage{xcolor}
\def\BibTeX{{\rm B\kern-.05em{\sc i\kern-.025em b}\kern-.08em
    T\kern-.1667em\lower.7ex\hbox{E}\kern-.125emX}}

\usepackage{titlesec}

\begin{document}

\title{On the Rate Region of I.I.D. Discrete Signaling and Treating Interference as Noise for the Gaussian Broadcast Channel
}

\author{
  \IEEEauthorblockN{Yujie Shao and Min Qiu}
  \IEEEauthorblockA{
                    Global College, Shanghai Jiao Tong University, Shanghai, China\\
                    Email: \{shaox3, min\_qiu\}@sjtu.edu.cn}
   \thanks{Corresponding author: Min Qiu}
}
\maketitle

\begin{abstract}
We revisit the Gaussian broadcast channel (GBC) and explore the rate region achieved by purely discrete inputs with treating interference as noise (TIN) decoding. Specifically, we introduce a simple scheme based on superposition coding with identically and independently distributed (i.i.d.) inputs drawn from discrete constellations, e.g., pulse amplitude modulations (PAM). Most importantly, we prove that the resulting achievable rate region under TIN decoding is within a constant gap to the capacity region of the GBC, where the gap is independent of all channel parameters. In addition, we show via simulation that the weak user can achieve a higher rate with PAM than with Gaussian signaling in some cases.
\end{abstract}

\begin{IEEEkeywords}
Gaussian broadcast channel, constellations, treating interference as noise.
\end{IEEEkeywords}

\section{Introduction}

Multiple access has been extensively studied in the literature~\cite{Clerckx2024}. 
Future multiple access systems are expected to provide higher system capacity and massive connectivity while accounting for fairness among users and devices~\cite{Jorswieck2024}. 
These future requirements highlight the critical need for communication strategies that can effectively manage interference and deliver robust performance when multiple users simultaneously share limited spectral resources.

When multiple users communicate simultaneously, signals intended for different users inevitably interfere with one another. Interference-management techniques such as successive interference cancellation (SIC) are commonly employed to remove multi-user interference. The idea of SIC can be traced back to the classical study of the Gaussian broadcast channel (GBC) and the Gaussian multiple access channel, where superposition coding of Gaussian signaling together with SIC decoding are the key ingredients to achieve the capacity~\cite{Tse_Viswanath_2005, Harshan2011, Harshan2013}. A large body of work has leveraged the fundamental results from the GBC in emerging communication applications, e.g.,~\cite{IslamNOMA2017,Clerckx2024,Jorswieck2024}. However, Gaussian signaling is difficult to realize in practice. Wireless communication systems typically adopt channel coding and discrete constellations, e.g., pulse amplitude modulation (PAM) and quadrature amplitude modulation (QAM). Consequently, design and analysis based on the Gaussian signaling assumption may not be directly applicable to practical communication systems~\cite{DeshpandeRajan2009}. Therefore, it is important to study communication schemes that employ realizable discrete signaling. 
Several works, e.g.,~\cite{Assaf2020, Bariah2019, Wang2017,Wei2020}, have studied the error probability of each user employing uncoded modulation (e.g., QAM) under SIC detection in the GBC and GMAC, respectively.
However, SIC introduces several challenges in the downlink model, where receivers are typically power and computational complexity-limited user equipment. The sequential nature of SIC leads to decoding complexity and latency that scales with the number of users, while its sensitivity to error propagation can severely degrade the performance of downstream users in the cancellation chain. Furthermore, SIC requires users to recover other users' messages, introducing potential privacy issues.

Compared with SIC, treating interference as noise (TIN) is more attractive in practice, as it only requires single-user decoding, resulting in single-user decoding complexity and latency. However, Gaussian signaling with TIN is strictly suboptimal when the interference is not very weak \cite{4675741}. This can be seen by noting that Gaussian is the worst noise (or interference when it is treated as noise) for the point-to-point channel~\cite{CoverThomas2006}. In contrast, discrete signaling can behave differently when treated as noise. In~\cite{DytsoTIN2016}, it has been proven that discrete signaling with TIN can achieve the optimal generalized degrees of freedom for the two-user Gaussian interference channel, thereby significantly outperforming Gaussian signaling with TIN. 
Motivated by the encouraging results in~\cite{DytsoTIN2016}, the authors in~\cite{ShiehHuang2016} designed a communication strategy by employing single-user coded PAM with TIN decoding and have proven that such a scheme is capable of achieving the capacity region of the $K$-user GBC to within a constant gap. The scheme was later generalized to a multi-dimensional lattice-partition multiple access framework~\cite{QiuTComm2018}, where the superimposed signaling still preserves the lattice structure, which can be exploited to harness inter-user interference in TIN decoding. It has been proven that a smaller gap to the GBC capacity is achieved when the base lattices are with larger shaping gains~\cite{QiuTComm2018}. The constant gap optimality of discrete signaling and TIN has also been shown to hold for the two-user Gaussian interference channel~\cite{Qiu2021}.

However,~\cite{DytsoTIN2016} uses a mixture of discrete and Gaussian inputs in some regimes, e.g., in the moderate weak interference regime, to balance the trade-off between achieving a high rate for the intended receiver and limiting the impact of interference at the non-intended receiver.
Moreover, the analytical gap to capacity therein is not always constant.
Although~\cite{ShiehHuang2016,QiuTComm2018,Qiu2021} consider purely discrete inputs, several issues still remain unresolved.
First, the constant-gap results were established only at a discrete set of rate points. In other words, it remains unclear whether the constant-gap result extends to the \emph{entire} achievable rate region with discrete signaling and TIN decoding, which itself has not yet been fully characterized.
In addition, the achievable schemes and proofs therein rely heavily on the linear deterministic model~\cite{Avestimehr2011} and the additional step by translating the schemes proposed for the deterministic model to the Gaussian model approximates the original Gaussian model and ignores the noise.
This two-step approach introduces approximation loss in the gap analysis and complicates the communication scheme design and proof.

In this paper, we take a different approach by designing and analyzing new achievable schemes based on discrete signaling with TIN \emph{directly} for the GBC without resorting to the linear deterministic model.
By characterizing a lower bound on each user's achievable rate, we prove that the \emph{whole} achievable rate region of the proposed scheme is within a constant gap to the whole capacity region of the GBC, where the gap is independent of channel parameters. 
In addition, we show via simulation that the weak user’s achievable rate with PAM can surpass that with Gaussian signaling in some cases.

\emph{Notations}: Random variables are written in upper-case sans serif font, e.g., $\msf{X}$. The floor and ceiling operations are denoted by $\lfloor \cdot \rfloor$ and $\lceil \cdot \rceil$, respectively.

\section{System Model}
\label{sec:system-model}

We consider a two-user real GBC with a single-antenna transmitter and two single-antenna receivers.
Let $\boldsymbol{x}_k$ denote a length-$n$ coded symbols intended for user $k\in\{1,2\}$.
The transmitter performs superposition coding via
\begin{equation}
    \boldsymbol{x} = \sqrt{P}\left(\sqrt{\alpha}\,\boldsymbol{x}_1 + \sqrt{1-\alpha}\,\boldsymbol{x}_2\right),
    \label{superposition}
\end{equation}
where $\alpha \in [0,1]$ is the power allocation factor for user~1 and $P$ is the total transmit power constraint such that $\frac{1}{n}\mathbb{E}\!\left[\|\boldsymbol{x}\|^2\right] \le P.$

The received signal at user~$k \in \{1,2\}$ is given by
\begin{equation}
  \boldsymbol{y}_k = h_k \boldsymbol{x} + \boldsymbol{z}_k,
  \label{eq:channel-model}
\end{equation}
where $h_k \in \mathbb{R}$ denotes the channel coefficient between the transmitter and user~$k$,
and $\boldsymbol{z}_k$ is additive white Gaussian noise with each element distributed over
$\mathcal{N}(0,1)$.
Without loss of generality, we assume $|h_1| > |h_2|$, so that user~1 is the strong user and user~2 is the weak user.
The signal-to-noise ratio for user~$k$ is defined as $\mathrm{SNR}_k \triangleq P|h_k|^2.$

The capacity region of two-user GBC is the collection of rate points $R_k\le C_k(\alpha)$ for $k \in \{1,2\}$ and $\alpha \in [0,1],$ where
\begin{align}
    &C_1(\alpha) = \frac12\log_2\!\left(1+\alpha\,\mathrm{SNR}_1\right),
    \label{C1}\\
    &C_2(\alpha)
  = \frac12\log_2\!\left(
 1 + \frac{(1-\alpha)\,\mathrm{SNR}_2}
          {1+\alpha\,\mathrm{SNR}_2}
 \right).
 \label{C2}
\end{align}

When SIC decoding is adopted, the strong user (user~1) first decodes $\boldsymbol{x}_2$ and subtracts it from $\boldsymbol{y}_1$ before decoding its own message.
In contrast, the weak user directly decodes its own message from $\boldsymbol{y}_2$ by treating the other user's signal as noise.
When TIN decoding is employed, each user directly decodes its own message from $\boldsymbol{y}_k$ by treating the other users' signals as noise.
In this work, we focus on TIN decoding due to its low complexity.

\section{Proposed Scheme And The Main Result}
\label{sec:achievable-scheme}

In this section, we introduce the proposed discrete signaling and the minimum distance analysis. We then present the main result of the paper.

\subsection{Proposed Signaling and Minimum Distance Analysis}
\label{subsec:pam}

In this section, we introduce the proposed purely discrete signaling.
For user $k\in \{1,2\}$, we let each element of $\boldsymbol{x}_k$ satisfy
$\msf{X}_k \sim \mathrm{PAM}\!\left(M_k,d_{\min}(\msf{X}_k)\right)$, meaning that $\msf{X}_k$ is uniformly distributed over a normalized PAM with zero mean, $M_k$ points, and minimum distance $d_{\min}(\msf{X}_k)=\sqrt{\frac{12}{M_k^2-1}}$.
The proposed scheme can employ other structured constellations such as QAM and multi-dimensional lattice constellations as in~\cite{QiuTComm2018}.

We let $N_k \triangleq \bigl\lceil \sqrt{\mathrm{SNR}_k}\,\bigr\rceil$,
which can be interpreted as a quantized effective channel-gain parameter.
We impose the following constraint on $(M_1,M_2)$:
\begin{equation}
  \prod\nolimits_{i=k}^{2} M_i \le N_k,\qquad k\in\{1,2\}.
  \label{eq:operating-region}
\end{equation}
Intuitively speaking, this constraint ensures the constellation size, and hence the number of transmitted bits remain within the channel’s supportable range.

Then, we analyze the minimum distance of the superimposed constellation, as it plays a fundamental
role in characterizing the achievable rate in the subsequent analysis.
We focus on a key power allocation regime for which the achievable rate is provably within a constant gap to capacity. In this regime, the constellation points of the superimposed constellation \emph{do not overlap}. The following lemma gives the minimum distance of
$\mathsf{X}=\sqrt{P}(\sqrt{\alpha}\msf{X}_1+\sqrt{1-\alpha}\msf{X}_2)$
under that power allocation regime.

\begin{figure}[t]
    \centering
    \includegraphics[width=0.95\columnwidth]{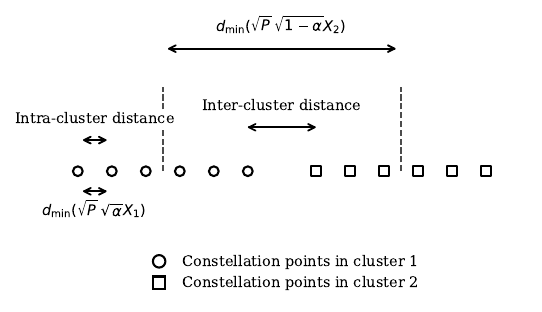}
    \caption{An example of the superimposed constellation in~\eqref{superposition} with $\msf{X}_1 \sim \mathrm{PAM}\!\left(6,d_{\min}(\msf{X}_1)\right)$ and $\msf{X}_2 \sim \mathrm{PAM}\!\left(2,d_{\min}(\msf{X}_2)\right)$. The intra-cluster and inter-cluster distances are indicated.}
    \label{fig1}
\end{figure}

\begin{lemma}
\label{lem:dmin}
Consider the superimposed constellation in~\eqref{superposition} with $\msf{X}_k$ uniformly distributed over a normalized zero mean $M_k\mathrm{-PAM}$ for $k\in\{1,2\}$.
If $0 < \alpha \le \alpha^\ast \triangleq \tfrac{M_1^2-1}{M_1^2M_2^2-1}$, then
\begin{equation}
d_{\min}(\msf{X})
=
\sqrt{\frac{12\alpha P}{M_1^2-1}}.
\end{equation}
\end{lemma}

\IEEEproof{
As shown in Fig.~\ref{fig1}, the inter-cluster distance is $\sqrt{\tfrac{12(1-\alpha) P}{M_2^2-1}}-(M_1-1)\sqrt{\tfrac{12\alpha P}{M_1^2-1}}$ and the intra-cluster distance is $\sqrt{\tfrac{12\alpha P}{M_1^2-1}}$ when $0 <\alpha \le \alpha^\ast$. As a result,
\begin{subequations}\label{eq:dmin}
\begin{align}
d_{\min}(\msf{X}) = &\min\!\left\{\sqrt{\tfrac{12(1-\alpha)P}{M_2^2-1}}-(M_1-1)\sqrt{\tfrac{12\alpha P}{M_1^2-1}},\,\sqrt{\tfrac{12\alpha P}{M_1^2-1}}\right\} \label{eq:dmin-a}\\
&=\sqrt{\tfrac{12\alpha P}{M_1^2-1}}. \label{eq:dmin-b}
\end{align}
\end{subequations}
}
\subsection{Main Result}
We first present a formal definition of within a constant gap notion. 

\begin{definition}\label{def1}
An achievable region is said to be \emph{within a constant gap $\Delta>0$} to the 
capacity region if for \emph{any} rate pair $(R_1, R_2)$ on the boundary of the achievable 
region, the rate pair $(R_1+\Delta, R_2+\Delta)$ is not achievable. Equivalently, $(R_1-\Delta,R_2-\Delta)$ 
is in the achievable region for any rate pair $(R_1,R_2)$ in the capacity region.\hfill$\blacksquare$

The main result of the paper is stated as follows.
\end{definition}
\begin{theorem}
\label{thm:main}
Consider the GBC in Section~\ref{sec:system-model},
the following rate region $\mathcal{R}$ is achievable with $\msf{X}_k$ uniformly distributed over a PAM constellation for $k \in \{1,2\}$ and TIN decoding. Specifically,
\begin{subequations}\label{R}
\begin{align}
&\mathcal{R}
\triangleq\;
\bigcup_{\substack{\alpha \in [0,\alpha^\ast]\\ \prod_{i=k}^{2} M_i \le N_k\\ k \in \{1,2\}}}
\Bigl\{
0 \le R_k \le I(\msf{X}_k;\msf{Y}_k)
\Bigr\}
\;\bigcup\;
\mathcal{R}_{\text{TS}},
\label{eq:R-a}
\\[1ex]
&\mathcal{R}_{\text{TS}}
\triangleq\;
\mathrm{co}\!\Biggl(
\bigcup_{\substack{\alpha \in \{\alpha^\ast,1\} \\
\prod_{i=k}^{2} M_i \le N_k\\ k \in \{1,2\}}}
\Bigl\{
0 \le R_k \le I(\msf{X}_k;\msf{Y}_k)
\Bigr\}
\Biggr),
\label{eq:R-b}
\end{align}
\end{subequations}
where $\mathrm{co}(\cdot)$ denotes the convex closure operation,
$\mathcal{R}_{\text{TS}}$ denotes the rate region obtained by time sharing (TS) between the achievable rate pairs under the power allocation $\alpha^\ast$ for all $(M_1, M_2)$ satisfying~\eqref{eq:operating-region} and user~1's single-user corner point,
and $\alpha^\ast$ is given in Lemma 1. 
Then, $\mathcal{R}$ is within a constant gap to the capacity region of the GBC in the sense of Definition \ref{def1}. \hfill$\blacksquare$
\end{theorem}
\begin{figure}[t]
    \centering
    \includegraphics[width=0.85\columnwidth]{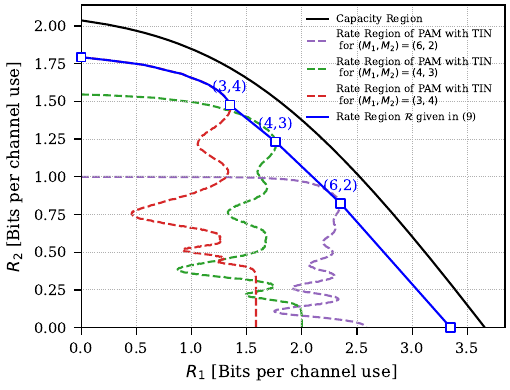}
    \caption{Achievable rate region of the proposed scheme for $(\mathrm{SNR}_1,\mathrm{SNR}_2)=(22,12)$ in dB. The rate pairs achieved with $\alpha^\ast$ are represented by blue boxes.}
    \label{fig3}
\end{figure}

\begin{remark}
From~\eqref{R}, the rate region $\mathcal{R}$ consists of two subregions. The first subregion is achieved by varying the power allocation between 0 and $\alpha^\ast$ for a certain pair of $(M_1,M_2)$ without TS. The second subregion is the TS rate region in~\eqref{eq:R-b}. We also emphasize that the proposed scheme not only provides greater flexibility in selecting the modulation order than \cite{ShiehHuang2016}, but also achieves a smaller gap to capacity in the case without TS. This improvement arises because our approach does not rely on the deterministic model. \hfill$\blacksquare$
\end{remark}
\begin{example}\label{ref:ex1}
Fig.~\ref{fig3} shows the rate region achieved by the proposed scheme and the capacity region. Observe that the rate region under $M_2=N_2$ and $\alpha \leq \alpha^\ast$ is close to the capacity region.
However, when $\alpha>\alpha^\ast$, the achievable rates for all $(M_1,M_2)$ (i.e., the dashed lines below the blue boxes) degrade severely. Hence, we use TS between the rate pairs achieved under larger $(M_1,M_2)$ and $\alpha^\ast$ to enlarge the rate region.\hfill$\blacksquare$
\end{example}
The proof of Theorem 1 is provided in Sec.~\ref{proof}.

\begin{remark}
As shown in Appendix B, 
the gap between the TS region of the two single-user capacity corner points and the capacity region can be arbitrarily large, highlighting the advantage of the proposed constant-gap-achieving scheme. \hfill$\blacksquare$
\end{remark}

\section{Proof of Theorem 1}
\label{proof}
Note that we omit the proof of $\alpha = 0$ and $\alpha = 1$ as they correspond to the single-user case, for which the achievable rate is within a constant gap (due to shaping loss) of the single-user capacity~\cite{DytsoTIN2016}.
\subsection{Achievable Rate Analysis}
We analyze the achievable rate of each user for $\alpha \in (0,\alpha^\ast]$. As in Example \ref{ref:ex1}, our scheme do not use $\alpha\in(\alpha^\ast,1)$.
\subsubsection*{1) Achievable Rate of User~1}
We bound user 1's mutual information under TIN decoding as follows.
\begin{subequations}\label{eq:MI_strong_lower_bound}
\begin{align}
&I(\msf{X}_1;\msf{Y}_1)
= I(\msf{X};\msf{Y}_1) - I(\msf{X}_2;\msf{Y}_1|\msf{X}_1)
\\
&\ge I(\sqrt{\mathrm{SNR}_1}\msf{X};\sqrt{\mathrm{SNR}_1}\msf{X}+ \msf{Z}_1) - H(\msf{X}_2)
\label{eq:MI_strong_lower_bound-a}\\
&\ge \log_2 M_1
-\frac12\log_2\!\left(\tfrac{2\pi e}{12}\right)
-\frac12\log_2\!\left(1+\tfrac{12}{d_{\min}(h_1\msf{X})^2}\right)
\label{eq:MI_strong_lower_bound-b}\\
& = \frac12\log_2\!\left(M_1^2+\alpha\mathrm{SNR}_1+ \frac{M_1^2-1}{\alpha\mathrm{SNR}_1}\right) \notag\\
& \quad- \log_2 M_1
 - \frac12\log_2\!\left(\tfrac{2\pi e}{12}\right),
\label{eq:MI_strong_lower_bound-c}
\end{align}
\end{subequations}
where~\eqref{eq:MI_strong_lower_bound-b} follows by applying Lemma 2 in Appendix to $I(\sqrt{\mathrm{SNR}_1}\msf{X};\sqrt{\mathrm{SNR}_1}\msf{X}+ \msf{Z}_1)$, and the resulting bound is simplified by the non-overlapping condition ensured by Lemma 1
for $\alpha \le \alpha^\ast$,~\eqref{eq:MI_strong_lower_bound-c} follows by applying Lemma 1 to $d_{\min}(h_1\msf{X})$.

\subsubsection*{2) Achievable Rate of User~2}
To bound user~2's mutual information under TIN decoding,
we define an effective noise $\msf{Z}_2'\triangleq \frac{\msf{Z}_2+ \sqrt{\alpha \mathrm{SNR}_2}\msf{X}_1}{1+\alpha\,\mathrm{SNR}_2}$ which has zero mean and satisfies
$\mathrm{Var}(\msf{Z}_2')=1$. Accordingly, we can construct an equivalent channel
$\msf{Y}_2'= \msf{X}_2'+\msf{Z}_2'$ from~\eqref{eq:channel-model} where $\msf{X}_2'=\frac{\sqrt{\mathrm{SNR}_2(1-\alpha)}\msf{X}_2}{\sqrt{1+\alpha\,\mathrm{SNR}_2}}$. As a result, user~2's mutual information is bounded as
\begin{align}
&I(\msf{X}_2;\msf{Y}_2) = I(\msf{X}_2';\msf{Y}_2')
\ge \log_2 M_2
-\frac12\log_2\!\left(\tfrac{2\pi e}{12}\right)
\notag\\
&\quad -\frac12\log_2\!\left(
1+\frac{1+\alpha\,\mathrm{SNR}_2}{(1-\alpha)\,\mathrm{SNR}_2}(M_2^2-1)
\right),
\label{eq:MI_weak_lower_bound}
\end{align}
where the last step follows from Lemma 2 in Appendix and the fact that $d_{\min}(\msf{X}_2')=\sqrt{\frac{12(1-\alpha)\mathrm{SNR}_2}{\sqrt{1+\alpha\,\mathrm{SNR}_2}(M_2^2-1)}}.$

\subsection{Gap Analysis for Power Allocation Regime $\alpha \in (0,\alpha^\ast]$}
We now analyze the gap between the achievable rate region and capacity region for $\alpha \in (0,\alpha^\ast]$. For this regime, we further consider two cases and define the gap as
\begin{equation}
\Delta_k(\alpha)
\triangleq
C_k(\alpha) - I(\msf{X}_k;\msf{Y}_k),\quad k \in \{1,2\}.
\label{delta}
\end{equation}
In order to approach the capacity region, we consider $M_1M_2=N_1$ to use the largest possible modulation size under the proposed constraint.

\textit{Case 1: $\alpha=\alpha^\ast$}: We are interested in achieving the largest rate pairs under the constraint in~\eqref{eq:operating-region}. Hence, we further place a constraint on~\eqref{eq:operating-region}. Since $M_2$ must be an integer, we let $M_2=\left\lfloor \frac{N_1}{M_1}\right\rfloor$
to fulfill the above constraint.

We first analyze the gap to capacity for user 1. Due to the additional constraint above, we have $(M_1M_2-1)^2<\mathrm{SNR}_1\le (M_1M_2+M_1)^2$ and $
\mathrm{SNR}_2>(M_2-1)^2.$
Let $\mathrm{SNR}_1=(M_1^2M_2^2-1)\delta$ for some constant $\delta$. It follows that
$    \tfrac{(M_1M_2-1)^2}{M_1^2M_2^2-1}<\delta \le \tfrac{(M_1M_2+M_1)^2}{M_1^2M_2^2-1}.$
By~\eqref{C1},~\eqref{eq:MI_strong_lower_bound} and using $\alpha^\ast$ from Lemma 1, we obtain that
\begin{subequations}\label{eq:gap1}
    \begin{align}
&\Delta_1(\alpha^\ast)
\le \frac12\log_2\!\left(
M_1^2
+\frac{M_1^2-1}{M_1^2M_2^2-1}\mathrm{SNR}_1 \right. \notag \\
&\quad\left.
+ \frac{M_1^2M_2^2-1}{\mathrm{SNR}_1}
\right)
- \log_2 M_1 + \frac12\log_2\!\left(\tfrac{2\pi e}{12}\right) \\
& = \frac{1}{2}\log_2\!\left(1+ \frac{M_1^2-1}{M_1^2}\delta + \frac{1}{M_1^2\delta} \right)
+ \frac12\log_2\!\left(\tfrac{2\pi e}{12}\right) \\
    & < \frac{1}{2}\log_2\!\left(1 + \frac{M_1^2-1}{M_1^2}\frac{(M_1M_2+M_1)^2}{M_1^2M_2^2-1} \right.\notag \\
    &\qquad \left.
    + \frac{1}{M_1^2}\frac{M_1^2M_2^2-1}{(M_1M_2+M_1)^2} \right) 
    + \frac12\log_2\!\left(\tfrac{2\pi e}{12}\right)
    \label{13c}\\
    & < \frac{1}{2}\log_2\!\left(1+\frac{12}{5}+\frac{1}{4} \right)
    + \frac12\log_2\!\left(\tfrac{2\pi e}{12}\right) =1.188,
    \label{13d}
\end{align}
\end{subequations}
where~\eqref{13c} uses the monotonicity over the range of $\delta$,~\eqref{13d} holds since we have $1<\frac{(M_1M_2+M_1)^2}{M_1^2M_2^2-1}\le\frac{12}{5}$ and $M_1\ge2,M_2 \ge2$.

To bound the gap to capacity for user 2, we use \eqref{C2}, \eqref{eq:MI_weak_lower_bound}, and $\alpha^\ast$ from Lemma 1, which gives
\begin{subequations}\label{user2gap}
    \begin{align}
    &\Delta_2(\alpha)
    \overset{\eqref{eq:MI_weak_lower_bound}}{\le} \frac{1}{2}\log_2\!\left(1 + \frac{(1-\alpha)\mathrm{SNR}_2}
    {1+\alpha\,\mathrm{SNR}_2}\right) - \log_2(M_2) \notag \\
    &\quad+ \frac{1}{2}\log_2\!\left(\tfrac{2\pi e}{12}\right)
    + \frac{1}{2}\log_2\!\left(
    1+\frac{1+\alpha\,\mathrm{SNR}_2}{(1-\alpha)\mathrm{SNR}_2}(M_2^2-1)
    \right)\label{www2}\\
    &= \frac{1}{2}\log_2\!\left(
    1
    + \frac{1}{M_2^2}\frac{(M_1^2M_2^2 - M_1^2)\,\mathrm{SNR}_2}
           {(M_1^2M_2^2-1) + (M_1^2-1)\mathrm{SNR}_2}
\right. \notag\\
&\qquad\left.
    + \frac{(M_1^2M_2^2-1)+(M_1^2-1)\mathrm{SNR}_2}
           {M_1^2M_2^2\mathrm{SNR}_2}
    \right)
+ \frac{1}{2}\log_2\!\left(\tfrac{2\pi e}{12}\right)  
\\
& < \frac{1}{2}\log_2\!\left(1 + \frac{M_1^2(M_2^2-1)}{M_1^2M_2^2-M_2^2} + \frac{1}{M_2^2}\frac{M_1^2M_2^2-1}{M_1^2\mathrm{SNR}_2}\right. \notag \\
&\qquad \left. + \frac{M_1^2-1}{M_1^2M_2^2}\right)
+\frac{1}{2}\log_2\!\left(\tfrac{2\pi e}{12}\right)  
\label{14c}\\
& < \frac{1}{2}\log_2\!\left(1+\frac{4}{3}+1+\frac{1}{4}\right) + \frac{1}{2}\log_2\!\left(\tfrac{2\pi e}{12}\right)= 1.175,
\label{14d}
\end{align}
\end{subequations}
where~\eqref{14c} follows since $M_1^2M_2^2>1$,~\eqref{14d} holds since we have $\frac{M_1^2(M_2^2-1)}{M_2^2(M_1^2-1)}<\frac{4}{3},\frac{1}{M_2^2}\le\frac{1}{4},\frac{M_1^2M_2^2-1}{M_1^2\mathrm{SNR}_2}<4, \frac{M_1^2-1}{M_1^2}<1$ when $\mathrm{SNR}_2> (M_2-1)^2$ and $M_1,M_2\ge2$.

\textit{Case 2: $\alpha\in(0,\alpha^\ast)$}:
In this regime, to approach the capacity region, we set $M_2=N_2$. 
Since $M_2$ must be an integer, we thus have $M_1=\left\lfloor \frac{N_1}{M_2}\right\rfloor$.

We start with user 1's gap analysis. We consider $\alpha\mathrm{SNR}_1>1$, since $\alpha\mathrm{SNR}_1 \le 1$ results in $C_1(\alpha)<1$, which means that the gap to capacity is already smaller than 1 bit/channel use.
Following~\eqref{eq:MI_strong_lower_bound}, we derive the gap to capacity for user~1 under $\alpha \in (0,\alpha^\ast)$ as 
\begin{subequations}
    \begin{align}
&\Delta_1(\alpha)
< \tfrac{1}{2}\log_2\!\left(M_1^2+\alpha\mathrm{SNR}_1
+ \frac{M_1^2-1}{\alpha\mathrm{SNR}_1}\right) \notag \\
&\quad - \log_2(M_1)
+ \tfrac{1}{2}\log_2\!\left(\tfrac{2\pi e}{12}\right) \notag\\
&<
\begin{cases}
\text{$\tfrac{1}{2}\log_2\!\left(2\right)
+ \tfrac{1}{2}\log_2\!\left(\tfrac{2\pi e}{12}\right)$}, & \text{C1}\\
\text{$\tfrac{1}{2}\log_2\!\left(\frac{2\pi e}{12}+\frac{2\pi e\alpha^\ast\mathrm{SNR}_1}{12M_1^2}
+ \frac{2\pi e(M_1^2-1)}{12\alpha^\ast M_1^2\mathrm{SNR}_1}\right)$}, & \text{C2}
\end{cases}
\label{www}\\
&<
\begin{cases}
    \text{$0.754$}, & \text{C1}\\
    \text{$1.188$}, & \text{C2}
\end{cases}
\end{align}
\end{subequations}
where~\eqref{www} follows that C1 represents the condition of $\mathrm{SNR}_1 < M_1^2M_2^2-1$ where the maximum is achieved at $\alpha=\frac{1}{\mathrm{SNR}_1}$ and C2 represents the condition of $\mathrm{SNR}_1 > M_1^2M_2^2-1$ where the maximum bound is achieved at $\alpha=\alpha^\ast$. 
Therefore, we conclude that for all $0<\alpha<\alpha^\ast$,
$\Delta_1(\alpha)<1.188.$

For user~2, we consider $\mathrm{SNR}_2>4$ because $\Delta_2(\alpha) < C_2(\alpha)<1.161$ for $\mathrm{SNR}_2\le4$.
Starting from~\eqref{www2}, we have
\begin{subequations}
    \begin{align}
&\Delta_2(\alpha)
\le \tfrac{1}{2}\log_2\!\left(M_2^2+\frac{(1-\alpha)\mathrm{SNR}_2}{1+\alpha\,\mathrm{SNR}_2}
\right.\notag \\
&\left. \quad+ \frac{1+\alpha\,\mathrm{SNR}_2}{(1-\alpha)\mathrm{SNR}_2}(M_2^2-1)\right)
-\log_2(M_2)
+\tfrac{1}{2}\log_2\!\left(\tfrac{2\pi e}{12}\right)\\
&< \tfrac{1}{2}\log_2\!\left(1+\frac{\mathrm{SNR}_2}{M_2^2}+\frac{M_2^2-1}{\mathrm{SNR}_2 M_2^2}\right)
 +\tfrac{1}{2}\log_2\!\left(\tfrac{2\pi e}{12}\right)\label{19c} \\
&< \tfrac{1}{2}\log_2(2+\tfrac{1}{M_2^2})
+\tfrac{1}{2}\log_2\!\left(\tfrac{2\pi e}{12}\right) \label{19d}\\
&< \tfrac{1}{2}\log_2\!\left(\tfrac{9}{4}\right)
+\tfrac{1}{2}\log_2\!\left(\tfrac{2\pi e}{12}\right)
=0.839, \label{19e}
\end{align}
\end{subequations}
where~\eqref{19c} follows that the maximum is achieved at $\alpha = 0$,~\eqref{19d} follows that $(M_2-1)^2<\mathrm{SNR}_2\le M_2^2$ and uses the monotonicity of the RHS of~\eqref{19c}, and~\eqref{19e} follows
because $M_2\ge2$.
Combining both cases, we conclude that for all $0<\alpha<\alpha^\ast$,
$\Delta_2(\alpha)< 1.661.$

\begin{figure}[t]
    \centering
    \includegraphics[width=0.85\columnwidth]{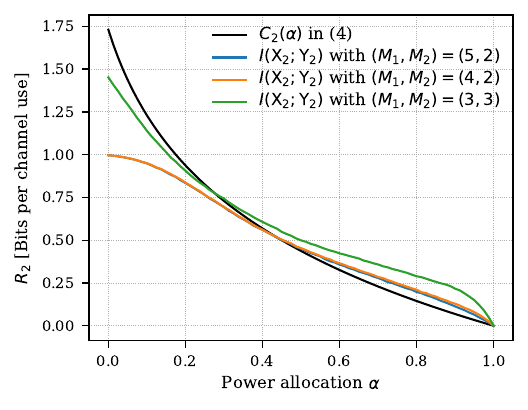}
    \caption{Comparison between user 2's achievable rate and $C_2(\alpha)$.}
    \label{fig5}
\end{figure}

\begin{example}
Fig.~\ref{fig5} shows $I(\msf{X}_2;\msf{Y}_2)$ evaluated by simulation, along with $C_2(\alpha)$ 
for $\alpha \in [0,1]$. We consider $(\mathrm{SNR}_1,\mathrm{SNR}_2)=(20,10)$ in dB, $(M_1,M_2)=(5,2),(4,2)$ and $(3,3)$.
Interestingly, $I(X_2;Y_2)$ outperforms $C_2$ in some regime. This may be because Gaussian signaling lacks structure and is the worst-case additive noise (or interference when treated as noise) in the point-to-point channel. In contrast, the structure of the interfering PAM signals can be exploited under TIN decoding.\hfill$\blacksquare$
\end{example}

\subsection{Gap Analysis for the Time Sharing Rate Region}
\label{sec:timesharing-alpha-star}

In this section, we analyze the gap between $\mathcal{R}_{\text{TS}}$ and the capacity region for $\alpha>\alpha^\ast$.
For the purpose of deriving the constant gap result, we focus on the rate region of TS among \emph{adjacent} boundary rate points achieved under the power allocation $\alpha^\ast$, which is a subset of $\mathcal{R}_{\text{TS}}$.

\begin{figure}[t]
    \centering
    \includegraphics[width=0.8\columnwidth]{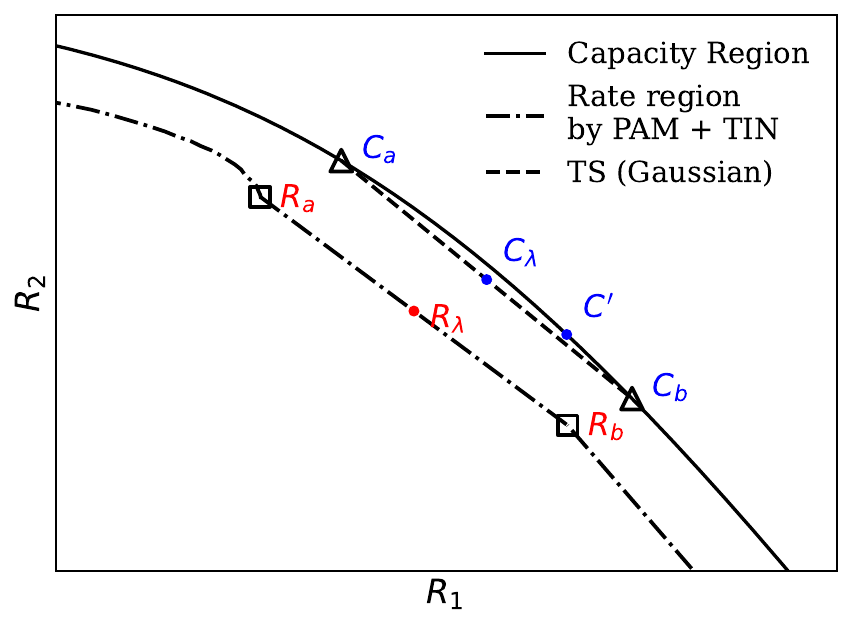}
    \caption{Illustration of achievable rate points.
}
    \label{fig2}
\end{figure}

Let $R_a\triangleq(R_{1,a},R_{2,a})$ and $R_b\triangleq(R_{1,b},R_{2,b})$
be two adjacent rate points achieved under the power allocation $\alpha^\ast$ 
and modulation orders $(M_1,\lfloor\frac{N_1}{M_1}\rfloor)$ and $(M_1+1,\lfloor\frac{N_1}{M_1+1}\rfloor).$ Let
$C_a\triangleq(C_{1,a},C_{2,a})$ and $C_b\triangleq(C_{1,b},C_{2,b})$ denote the corresponding Gaussian boundary points where $C_{1,b}>C_{1,a}$ and $C_{2,b}< C_{2,a}$.
Moreover, any rate point on the TS region between $R_a$ and $R_b$ can be represented by $R_{\lambda} = (R_{1,\lambda},R_{2,\lambda})$ where $R_{\lambda}=\lambda R_a+(1-\lambda)R_b$ and $\lambda\in[0,1]$ is the TS parameter. Similarly, we introduce the TS point $C_{\lambda}=(C_{1,\lambda},C_{2,\lambda})$ between $C_a$ and $C_b$.

Since both $R_{\lambda}$ and $C_{\lambda}$ are convex combinations of their respective
endpoints, the same constant-gap bounds in \eqref{eq:gap1} and \eqref{user2gap} hold for all $\lambda$ such that $C_{1,\lambda}-R_{1,\lambda}<1.188$ and $
C_{2,\lambda}-R_{2,\lambda}<1.175.$

Let $C'=(C_{1}',C_{2}')$ be any Gaussian boundary point on the curve between $C_a$ and $C_b$.
An illustration of the above points is given in Fig.~\ref{fig2}. Our approach to the proof consists of two steps. First, we upper-bound the gap between $C'$ and $C_\lambda$.
Second, using the triangle inequality, we further bound the gap between $C'$ and $R_\lambda$.

In the first step, 
since $C_{1}'$ is between $C_{1,a}$ and $C_{1,b}$, the gap between $C'$ and $C_{\lambda}$ can be upper bounded  by
\begin{subequations}
    \begin{align}
 &C_{1}'-C_{1,\lambda}
 < C_{1,b}-C_{1,a}  \\
 & < \frac{1}{2}\log_2\!\left(\frac{(N_1-M_1-1)^2-1+\mathrm{SNR}_1((M_1+1)^2-1)}{N_1^2-1+\mathrm{SNR}_1(M_1^2-1)} \right) \notag \\ 
 &\quad +  \frac{1}{2}\log_2\!\left( \frac{N_1^2-1}{(N_1-M_1-1)^2-1}\right) \label{21b}\\
 & < \tfrac{1}{2}\log_2\!\left(\frac{(M_1+1)^2-1}{M_1^2-1}\,\frac{N_1^2-1}{(N_1-M_1-1)^2-1} \right)   \label{21c}\\
 & < \tfrac{1}{2}\log_2\!\left(\frac{8}{3} \right) + \tfrac{1}{2}\log_2\!\left(\frac{35}{8} \right) = 1.772,\label{21d}
\end{align}
\end{subequations}
where~\eqref{21b} follows that $x-1<\lfloor x \rfloor \le x$ for any real $x$,~\eqref{21c} uses monotonicity of $\tfrac{(N_1-M_1-1)^2-1+\mathrm{SNR}_1((M_1+1)^2-1)}{N_1^2-1+\mathrm{SNR}_1(M_1^2-1)}$ over $\mathrm{SNR}_1$,~\eqref{21d} follows that the maximum value of the function is achieved taking the partial derivative for $M_1\ge2$ and $N_1\ge2(M_1+1)$.

Similarly, the gap between $C'$ and $C_\lambda$ for user 2 can be derived as follows.
\begin{subequations}
    \begin{align}
 &C_{2}'-C_{2,\lambda}
 < C_{2,a}-C_{2,b}  \\
 &< \frac{1}{2}\log_2\!\left( \frac{N_1^2-1}{(N_1-M_1-1)^2-1}\right) \notag \\
 &\quad + \frac{1}{2}\log_2\!\left(\frac{N_1^2-1+\mathrm{SNR}_2((M_1+1)^2-1)}{(N_1-M_1)^2-1+\mathrm{SNR}_2(M_1^2-1)} \right) \label{22b}\\
 & < \frac{1}{2}\log_2\!\left( \frac{N_1^2}{(N_1-M_1-1)^2}\right) \notag \\
 & \quad+\frac{1}{2}\log_2\!\left(\frac{N_1^2-1+\mathrm{SNR}_2((M_1+1)^2-1)}{\mathrm{SNR}_2(M_1^2-1)} \right) \label{22c} \\
 &< \frac{1}{2} \log_2\!\left(\frac{35}{8} \right) + \frac{1}{2} \log_2\!\left(\frac{35}{3} + \frac{8}{3} \right) = 2.985,\label{22d}
\end{align}
\end{subequations}
where~\eqref{22b} follows that $x-1<\lfloor x\rfloor\le x$ for any real $x$,~\eqref{22c} follows since $(N_1-M_1)^2-1>0$ and~\eqref{22d} follows that $ \frac{N_1^2-1}{(N_1-M_1-1)^2-1}\le \frac{35}{8}$ for $N_1\ge2(M_1+1)$ and $\frac{N_1^2-1}{\mathrm{SNR}_2(M_1^2-1)}\le \frac{35}{3}$ and $\frac{(M_1+1)^2-1}{M_1^2-1}\le\frac{8}{3}$ for $M_1\ge2$ and $ N_1\ge2(M_1+1).$

Using the triangle inequality, we obtain the gap between the TS rate region and the capacity region as
\begin{align}
 C_{1}'-R_{1,\lambda}
 &< |C_{1}'-C_{1,\lambda}| + |C_{1,\lambda}-R_{1,\lambda}|
 < 2.960 , \\
 C_{2}'-R_{2,\lambda}
 &< |C_{2}'-C_{2,\lambda}| + |C_{2,\lambda}-R_{2,\lambda}|
 < 4.160.
\end{align}

Due to space limitations, we omit the gap analysis for the rate region achieved by TS between user 1's single-user corner point and its adjacent rate point achieved with $\alpha^\ast$, as the proof follows the same line of approach above and the constant gap result holds.

\section{Conclusion}
In this paper, we revisited the GBC and introduced a new achievable scheme based on discrete signaling and TIN decoding. Unlike previous works, we directly designed and analyzed our scheme for the GBC without using the linear deterministic model, thereby leading to simpler designs and proofs.
We have proven that the \emph{whole} rate region achieved by the proposed scheme
is within a constant gap to the capacity region. 
In addition, we showed that the weak user's rate achieved by PAM can be larger than that by Gaussian signaling in some cases.
Our results imply that practical discrete signaling 
and TIN decoding are promising approaches for interference management and achieving near-capacity performance.
As part of future work, we plan to extend the proposed design and analysis to the GBC with more than two users.

\section*{Appendix A}
\begin{lemma}[Proposition~1 of~\cite{DytsoTIN2016}]
Let $\msf{F}$ be a discrete random variable with minimum Euclidean distance
$d_{\min}(\msf{F})>0$, and let $\msf{Z}$ be a zero-mean, unit-variance random variable
independent of~$\msf{F}$. Then
\begin{equation}
\begin{aligned}
I(\msf{F};\msf{F}+\msf{Z})
\ge& H(\msf{F})
- \frac12\log_2\!\left(\tfrac{2\pi e}{12}\right) \notag \\
& - \frac12\log_2\!\left(1+\tfrac{12}{d_{\min}(\msf{F})^2}\right).
\label{eq:lemma_MI_bound}
\end{aligned}
\tag{20}
\end{equation}
\end{lemma}

\section*{Appendix B}
We show that the gap between the TS region of the two single-user capacity corner points and the capacity region can be arbitrarily large. To justify this, we use, following~\cite{erez2017commentsdownlinknonorthogonalmultiple}, the relative gain defined at the sum-rate-optimal point, which is the common corner point at which TS and Capacity coincide. By perturbing slightly away from this point, one can directly compare how efficiently the two schemes transfer rate from the stronger user to the weaker user, and this is captured by the local slope and the resulting relative gain. Specifically,
\begin{align}
&g(\mathrm{SNR}_1,\mathrm{SNR}_2)
= \frac{\text{slope}_{\text{Cap}}}{\text{slope}_{\text{TS}}}\notag\\
&\quad= \frac{\mathrm{SNR}_2(1+\mathrm{SNR}_1)}{\mathrm{SNR}_1(1+\mathrm{SNR}_2)}
\frac{\log_2(1+\mathrm{SNR}_1)}{\log_2(1+\mathrm{SNR}_2)},
\end{align}
where $\text{slope}_{\text{TS}}$ and $\text{slope}_{\text{Cap}}$ denote the slopes of the capacity boundary and the TS boundary, evaluated at the sum-rate-optimal point, and are given in (6) and (7) of~\cite{erez2017commentsdownlinknonorthogonalmultiple}, respectively. It is not difficult to see that $g(\mathrm{SNR}_1,\mathrm{SNR}_2)$ can become arbitrarily large for some SNR pairs. Hence, the gap between the TS region and the capacity region can also be arbitrarily large.

\bibliographystyle{IEEEtran}
\bibliography{refs}

\end{document}